\documentclass{ws-ijgmmp}

\begin{document}

\markboth{Hermano Velten}
{Viscous Cold Dark Matter in agreement with observations}

%
\catchline{}{}{}{}{}
%

\title{Viscous Cold Dark Matter in agreement with observations}

\author{Hermano Velten}

\address{Departamento de F\'isica, CCE, Universidade Federal do Esp\'irito Santo,\\ Campus Goiabeiras 514, 29075-910, Vit\'oria, ES, Brazil\, \\
and  \\
Fakult\"at f\"ur Physik, Universit\"at Bielefeld, Universit\"atstra\ss e 25, \\
Bielefeld, 33501, Germany\, \\
\email{velten@physik.uni-bielefeld.de}}

\maketitle

\begin{history}
\received{(Day Month Year)}
\revised{(Day Month Year)}
\end{history}

\begin{abstract}
We discuss bulk viscous cosmological models. Since the bulk viscous pressure is negative, viable viscous cosmological scenarios with late time accelerated expansion can in principle be constructed. After discussing some alternative models based on bulk viscous effects we will focus on a model very similar to the standard $\Lambda$CDM. We argue that a $\Lambda${\rm v}CDM model, where we assign a very small (albeit perceptible) bulk viscosity to dark matter is in agreement with available cosmological observations. Hence, we work with the concept of viscous Cold Dark Matter ({\rm v}CDM). At the level of the perturbations, the growth of {\rm v}CDM structures is slightly suppressed when compared with the standard CDM ones. Having in mind that the small scale problems of the $\Lambda$CDM model are related to an excess of clustering, our proposal seems to indicate a possible direction for solving the serious drawbacks of the CDM paradigm within the standard cosmological model. 

\end{abstract}


\section{Introduction}

It is evident for any cosmologist the analogy between cosmological models and hydrodynamics. In this framework, cosmic components are usually modeled as ideal and perfect fluids, but in the real universe many dissipative processes can take place. For instance, particle production in time-dependent gravitational fields \cite{partprod}, diffusion phenomena \cite{calogero} and viscosities are, eventually, likely to occur during the cosmic evolution. 

The aim of this work is to discuss cosmology beyond the perfect fluid behavior established by the standard model. In order to make such transition from the standard approach to the dissipative one, let us first introduce the dynamics of the so called $\Lambda$CDM model.

The standard description for the homogeneous and isotropic expansion of the universe, i.e. a Friedmann-Robertson-Walker (FRW) metric, can be easily simplified by the expression
\begin{equation}\label{Hz}
H^2 (a)=H^2_0 \left[\frac{\Omega_{\rm r0}}{a^4}+\frac{\Omega_{\rm m0}}{a^3}+\frac{\Omega_k}{a^2}+\Omega_{\Lambda}\right],
\end{equation}
where $H(a)=\dot{a}(t)/a(t)$. The scale factor today is normalized as $a_0=1$. The today's value for the Hubble parameter $H_0$ is constrained by different observations. The PLANCK satellite has provided the most recent result, $H_0$=67.3 Km s$^{-1}$Mpc$^{-1}$ \cite{Planck2013CP}, which is the lowest value found in the literature. Another common value is $H_0$=72.0 Km s$^{-1}$Mpc$^{-1}$ from the Hubble Space Telescope $(HST)$ \cite{HST}, but some estimations can reach up to $H_0$=74.8 Km s$^{-1}$Mpc$^{-1}$ \cite{Riess} and even $H_0$=78.7 Km s$^{-1}$Mpc$^{-1}$ \cite{Suyu}. The fractionary densities are defined as $\Omega = \rho / \rho_c$, where $\rho_c$ is the critical density. For relativistic components $\Omega_{\rm r0}$ is decomposed into photons $\Omega_{\rm \gamma 0}$ and neutrinos $\Omega_{\rm \nu 0}$, ($\Omega_{\rm r0}=\Omega_{\rm \gamma 0}+ \Omega_{\nu 0})$. The relativistic pressure $p=\rho/3$ originates the scaling relation $\Omega_{\rm \gamma} \propto a^{-4}$. In fact, the contribution of massive neutrinos is assumed to be an extension of the standard model. The mass of neutrinos determines the transition to the non-relativistic phase. Then, for the late universe neutrinos can potentially play the role of matter. But, it is usual to neglect the neutrino background $\Omega_{\nu 0}=0$. The matter components are composed by baryons and cold dark matter $(\Omega_{\rm m}=\Omega_{\rm b} + \Omega_{\rm cdm})$. The pressure $p_{\rm m}=0$ produces the scaling $\propto a^{-3}$ in Eq. (\ref{Hz}). A curvature contribution $\Omega_{k}$ is, in principle, also allowed in a FRW metric. And, finally, a dark energy component, here in the form of a cosmological constant $\Omega_{\Lambda}$, is an essential ingredient in order to produce the late time accelerated expansion.

It is worth noting that such simple formula fits all the observed distance measurements. Supernovae data \cite{SN}, Baryonic acoustic oscillations \cite{bao}, indirect estimation for $H(z)$ \cite{Hz} and other probes are well fitted by the $\Lambda$CDM cosmology. It is also remarkable that within the $\Lambda$CDM free parameter space $\left\{H_0, \Omega_{\rm \gamma0}, \Omega_{\rm m0}, \Omega_{\rm k}, \Omega_{\rm \Lambda} \right\}$, only one specific point is capable to describe, at the same time, all the main cosmological observations with a high statistical confidence. In practice, we can summarize the best-fit $\Lambda$CDM concordance model with the following set of values
\begin{equation}
H_0=70.0 ~{\rm Km \,s}^{-1}{\rm Mpc}^{-1};~ \Omega_{\rm r0}\approx10^{-5};~ \Omega_{\rm m0}\cong 0.3;~ \Omega_{\rm k}\approx 0;~ \Omega_{\rm \Lambda}\cong 0.7,
\end{equation}
remember that the $\Omega_{b0}$ is well constrained be the nucleosynthesis. In practice, a flat cosmology is usually adopted ($\Omega_{k}=0$) and the radiation contribution is negligible for the late time cosmic dynamics.

Within the standard $\Lambda$CDM model dissipative effects are usually neglected. All components of such model are modeled as perfect (adiabatic) ones. To exemplify what does this mean, we have to discuss how density perturbations evolve in such model. In cosmological perturbation theory we decompose the energy density of the fluid $\rho$ as $\rho=\bar{\rho}+\hat{\rho}$, where the symbol bar denotes a background quantity - that enters in Eq. (\ref{Hz}) - and the hat means a first order perturbation. Applying the same to the pressure $P$ of the fluid we have $P=\bar{P}+\hat{P}$. Adiabatic perturbations are characterized by
\begin{equation}
\hat{P}_{eff}-\frac{\dot{\bar{P}}_{eff}}{\dot{\bar{\rho}}}\hat{\rho}=0.
\end{equation}
It is possible to identify $c^2_{ad}=\dot{P}_{eff}/\dot{\rho}$ as the adiabatic speed of sound. In other words, in adiabatic models, the total or effective speed of sound is equal to the adiabatic one.

One can argue that in a real universe it is very unlikely that dissipative processes do not take place. Particle production is, in principle, allowed in a expanding space-time. However, there are no direct observation that such process happens in nature. Another possible dissipative phenomena is diffusion, but its inclusion in the context of general relativity can lead to some difficulties in preserving the Bianchi identities. Concerning the possible viscosities that can take place in the universe, shear viscosity and heat conduction, for example, are directional processes and the cosmological principle impedes their existence in a FRW metric. Hence, bulk viscosity is the unique process allowed in a expanding, homogeneous and isotropic background.

The main difference between the standard model and other dissipative approaches lies on the behavior of the cosmological perturbations. While perturbations in the standard cosmology are always adiabatic, dissipative models of the dark sector are intrinsically non-adiabatic. Such feature can be seen in the perturbations if  
\begin{equation}
\hat{P}_{eff}-\frac{\dot{\bar{P}}_{eff}}{\dot{\bar{\rho}}}\hat{\rho}\neq0.
\end{equation}

The above relation allow us to define the effective speed of sound $c^2_{eff}$ in terms of $c^2_a$ and the viscous contribution $c^2_{vis}$ as
\begin{equation}
c^2_{eff}=c^2_{ad}+c^2_{vis}.
\end{equation}

As we show bellow, bulk viscosity produces a very specific type of non-adiabatic perturbations. In principle, high values of $c^2_{vis}$ are not allowed by large scale structure observations. However, our aim in this work is to discuss how relativistic bulk viscous fluids, i.e., even allowing the existence of non-adiabatic perturbations, can be used to construct viable cosmological scenarios. In particular, and perharps the most interesting application, we show that bulk viscosity can be accommodated within the standard cosmological scenario. We focus on a scenario where the dark matter component of the $\Lambda$CDM model has a bulk viscous pressure. We also argue that for any viscous model, the analysis of the structure formation process places the strongest constraints on the viscosity of cosmic fluids due to the non-adiabatic nature of the viscous perturbations.

\section{Bulk viscous effects in cosmology}

A fluid description in terms of the bulk viscous properties has been widely applied to cosmology since the development of relativistic thermodynamics. Bulk viscosity, which is also known as second viscosity \cite{landau} (in this nomenclature, the first viscosity is the shear) is associated to a non-equilibrium pressure (or, the dynamic pressure). However, it is worth noting the in non-equilibrium thermodynamics the viscous pressure is a small correction to the equilibrium pressure. Then, it is important to keep in mind that the possible viscous effects in cosmology can not play a decisive role for the total dynamics. This is true for theories that take into account both first and second order deviations from the equilibrium. 
However, we can wonder whether a small (and allowed) viscosity is able to leave some imprint on the cosmic evolution. If the effects of bulk viscosity really exist in the universe we have to be able to indirect observe them with help of astronomical observations.

The first order theory, which can also be called Eckart's theory \cite{eckart} (or even Landau's theory \cite{landau}), represents the traditional way to study bulk viscosity. It has been developed during the 1940s and 1950s. However, it became clear during the late 1960s and 1970s that this type of approach suffers from causality and stability problems. The inclusion of the second order deviations from equilibrium became mandatory in order to solve such problems. This led to the second order or the M\"uller-Israel-Stewart theory (MIS) \cite{muller, Israel, IStewart}. See also \cite{diego, hiscock}. 

The second order theory introduces a new parameter which is the relaxation time. Apart from this, the theory itself is more complicated and deserves a more careful physical interpretation of the relevant quantities than the first order one. This explains why it is so rare to find cosmological applications of the second order theory. On the other hand, the Eckart theory is widely studied and one can find in the literature many works on cosmology based on this approach. We will also develop this work using the Eckart frame. However, in the last section we turn our attention to the MIS theory and will discuss some relevant aspect of this theory to cosmology.

In the Eckart frame the first order deviations from equilibrium are expressed as additional contributions to the energy-momentum tensor $\Delta  T^{\mu \nu} $ as
\begin{equation}
T^{\mu \nu}= \rho u^{\mu}u^{\nu}+p_{k} \, h^{\mu\nu} +\Delta T^{\mu \nu}, \hspace{0.5cm} {\rm with} \hspace{0.5cm} \Delta T^{\mu \nu}=-\xi u^{\gamma}_{; \gamma} h^{\mu\nu}.   
\end{equation}
We defined $ h^{\mu\nu}=u^{\mu}u^{\nu}+g^{\mu\nu}$. The coefficient of bulk viscosity $\xi$ is positive due to the second thermodynamics law \cite{weinberg1}. For the background, $u^{\gamma}_{; \gamma}=3H$ which means that bulk viscosity modifies the effective pressure as
\begin{equation}
P_{eff}=p_{k}-3H\xi,
\end{equation}
where $p_{k}$ is the kinetic pressure. The quantity $\Delta T^{\mu \nu}$, which can also include shear viscosity and heat conduction contributions, is constructed in such way that the conservation ($T^{\mu}_{\nu \, ; \; \mu}=0$) still holds in the presence of dissipative contributions. 

A standard assumption in cosmology is that the conservation of $T^{\mu \nu}$ holds separately for each cosmic component. For a typical equation of state $P_{eff}=w\rho$ one finds
\begin{equation}\label{conserv}
\dot{\rho}+3H\rho(1+w)=0.
\end{equation}
For adiabatic ($\xi=0$) cosmic fluids the equation of state parameter assumes the values $w_{\rm m}=0$ for matter (dark matter and baryons) and $w_{\rm r}=1/3$ for relativistic (neutrinos and photons). Solving (\ref{conserv}) for these fluids we find $\rho_{\rm m}\propto (1+z)^3$ and $\rho_{\rm r}\propto (1+z)^4$, respectively. Dark energy can also be described by $P_{\rm de}=w_{\rm de}\rho_{\rm de}$ with $w_{\rm de} <-1/3$, where the cosmological constant $\Lambda$ is recovered if $w_{\rm de}=-1$. Hence, it is obvious that if bulk viscosity is allowed the background dynamics is somehow modified because the density evolution of such viscous fluid will be different. On has to solve Eq. (\ref{conserv}) with the appropriate form of $w$ which includes the viscous pressure. It is also expedient to note that bulk viscosity is the unique effect in nature which is able to reduce the kinetic pressure of a fluid.

Let us now just make a brief historical review of some of the most remarkable cosmological applications of viscous imperfect fluids. The first works on a possible cosmological bulk viscosity appeared in the 1970s \cite{IsraelV, Klimek, Murphy, Belinskii, weinberg}. The first applications concerned the early time cosmology. Indeed, a bulk viscous pressure in the early
universe can be the result of cosmological particle production \cite{Zel, barrow}. Due to the fact that bulk viscous pressure is negative, an inflationary epoch driven by bulk viscous pressure has also been studied in the 1980s \cite{Diosi, waga, maartens95, winfried96, maartens97, winfried00}. All these works have analysed the role played by bulk viscosity in the early universe. However, much before the discovery of the accelerated expansion of the universe (the dark energy phenomena) in 1998, one can find some mentions for a late time viscous universe \cite{potuba, padmanabhan}. The late accelerated universe as an effect of the bulk viscosity in the cosmic media has been first investigated in refs. \cite{zimdahl, balakin}. 

In general, all these applications rely on the phenomenological ground and are just assumptions based on the possible existence of a cosmic bulk viscosity. However, there are some attempts in the literature to justify the cosmological bulk viscosity \cite{winfried1}. It is not clear which cosmic fluid has such bulk viscosity, but it has been demonstrated a long time ago that a gas of neutrinos have bulk viscous properties \cite{degroot} and it is quite surprising that analysis in the field of neutrino cosmology do not take neutrino bulk viscosity into account.

\section{Do we really need dark energy?}

In the last section we have shown that bulk viscosity is able to induce a negative pressure. It is not expected for ordinary fluids to display negative pressure in experiments. However, cosmologists known that the dark energy phenomena can be explained only via the inclusion of such exotic fluids with negative equation of state parameters. Then, a first approach for using bulk viscosity in cosmology seems to be the use of such fluid as a dark energy candidate \cite{viscousDE, darkgoo}.

Hence, assuming that bulk viscous pressure is present in either the dark or the baryonic matter distributions, do we really need dark energy in order to explain the accelerated expansion? As the primordial nucleosynthesis sets that the abundance of baryonic matter is only of order $\Omega_{\rm b0}\sim 0.045$, only a huge (and therefore unlikely) viscosity in the baryonic sector would serve for the purpose of accelerating the universe. However, remembering our ignorance about the dark sector of the universe and thus assuming that only dark matter is a bulk viscous fluid, is the negative bulk viscous pressure of dark matter able to drive the accelerated expansion? As we will argue in this section, the answer is positive. Let us now describe the background expansion of a viable cosmology without dark energy.

We write the Hubble expansion as
\begin{equation}\label{Hzviscous}
H^2(a)=H^2_0 \left[\frac{\Omega_{\rm r0}}{a^4}+\frac{\Omega_{\rm b0}}{a^3}+\Omega_{\rm vm}(a)+\frac{\Omega_k}{a^2}\right].
\end{equation}

The above expansion (\ref{Hzviscous}) is known as unified (or ``quartessence'') model, in the sense that dark matter and dark energy are seem as a unique substance. Then, if a flat $\Omega_{k}=0$ cosmology is adopted, we have $\Omega_{\rm vm0}=1-\Omega_{\rm b0}-\Omega_{\rm r0}\approx 0.95$. The viscous dark matter $\Omega_{\rm vm}$ has to describe both the dark matter and dark energy properties simultaneously. It is a function that interpolates from the typical CDM behavior $\Omega_{\rm vm}(z>>0)\propto (1+z)^3$ in the past, where structures form, to a dark energy form with $\Omega_{\rm vm} (z\sim 0) = const$ for recent times. The idea of unification of the dark sector was first applied to cosmology using the Chaplygin gas as the candidate for the unified fluid \cite{chaplygin}. But, the use of a bulk viscous for the unification scheme was proposed almost at the same time in Ref. \cite{schwarz02}.

In order to describe its dynamics in this case, let us set the pressure (with $p_{k}=0$) as
\begin{equation}\label{Pvis}
P_{\rm vm}=-\xi u^{\gamma}_{;\gamma}=-3H \xi.
\end{equation}
The main aspect of any viscous fluid is the coefficient $\xi$. From the relativistic kinetic theory we known that the bulk viscosity is a transport coefficient proportional to the temperature as $\xi\propto T^{m}$, where $m$ is a positive quantity \cite{kremer}. Here, since we are dealing with a fluid description, we adopt
\begin{equation}\label{xi}
\xi=\xi_0\left(\frac{\rho_{\rm vm}}{\rho_{\rm vm0}}\right)^{\nu}.
\end{equation}
Thus, a theoretical prior on the exponent $\nu$ seems to be $\nu>0$. However, most of the applications do not take this into account and allow $\nu$ to assume negative values.

The bulk viscous fluid has been widely used as a candidate for the unified model \cite{unified}. In order to briefly demmonstrate how a bulk viscous fluid provides an unified scenario, let us neglect in a first moment the contribution of baryons and radiation. Then, $H\sim \rho_{\rm vm}^{1/2}$. The pressure of the fluid can be written as $P_{\rm vm}\sim -H\xi\sim-H \rho_{\rm vm}^{\nu}\sim-\rho_{\rm vm}^{\nu+1/2}$. Inserting this pressure into Eq. (\ref{conserv}) we find
\begin{equation}
\rho_{\rm vm}=\left(\frac{3H_0\xi_0}{\rho_{\rm vm0}}+\frac{1-\frac{9H_0\xi_0}{\rho_{\rm vm0}}}{a^{3\left(\frac{1}{2}-\nu\right)}}\right)^{\frac{1}{\frac{1}{2}-\nu}}.
\end{equation}

The existence of an early matter dominated epoch, $H(a<<1) \sim a^{-3/2}$, is
guaranteed for $\nu < 1/2$ and $\xi_0 < \rho_{0}/(3H_0)$.

The unified bulk viscous fluid shows competitive results at background level and even concerning the matter power spectrum data \cite{unified}. However, as pointed out in Ref. \cite{LiBarrow}, the viscous unified model is not compatible with the CMB data. The general festure observed is a huge power on large scale, see Fig. 5 in \cite{LiBarrow}. This is caused by an increase in the integrated Sachs-Wolfe (ISW) signal which is proportional to the time derivative of the gravitational potential at large scales. This means that the nonadiabatic perturbations can significantly modify the first order dynamics. In order to visualise the source of this apparent problem, let us assume a line element for scalar perturbations in the Newtonian gauge without anisotropic stress
\begin{eqnarray}\label{Newmetric}
ds^{2}=a^{2}\left(\eta\right)\left[-\left(1+2\psi\right)d\eta^{2}+\left(1-2\psi\right)\delta_{ij}dx^{i}dx^{j}\right],
\end{eqnarray}
where $\eta$ is the conformal time. If we compare the perturbations of the pressure in the adiabatic case with the perturbations of the bulk viscous pressure (the nonadiabatic situation) we have
\begin{eqnarray}
\label{pad}P_{ad}\equiv P_{ad}(\rho) \rightarrow \delta P_{ad}&=& c^2_{ad} \rho \Delta, \\
\label{pnad}P_{\rm vm}\equiv-\xi u^{\mu}_{; \mu} \rightarrow \delta P_{\rm vm}&=&-3H \delta\xi -\xi 
\delta(u^{\mu}_{; \mu}) \\ \nonumber
&=&-3H\xi\nu\Delta-\xi\left(\delta u^i_{,i}-\frac{3\mathcal{H}\psi}{a}-\frac{3\psi^{\prime}}{a}\right)
\end{eqnarray}
where $\Delta=\hat{\rho}/\bar{\rho}$ is the density contrast, the symbol prime means derivative with respect to $\eta$ and $\mathcal{H}=a^{\prime}/a$. Note that in order to write $\delta P_{\rm vm}$ in terms of $\Delta$ only, it is necessary to use the perturbed $0-0$ and $0-i$ components of the Einstein equation. The gravitational potential can be calculated directly with use of the $i-j$ component and it has the perturbation $\delta P$ is the source term. Then, by a simple comparison of relations (\ref{pnad}) are (\ref{pnad}) one sees why the gravitational potential changes from the adiabatic to the nonadiabatic case.

Indeed, the ISW effect is a very sensitive probe for viscous models. However, it has been shown in Ref. \cite{VeltenSchwarz11} that with a proper choice of the parameters of the unified viscous model one can explain the CMB data. However, in this same reference, it has been pointed out that an analysis of the growth of viscous dark matter halos, through a kind of ``viscous Meszaros equation'', is able to place very strong constraints on the viscous dark fluid. It is also shown that the source of the difficulties that viscous models have faced is related the contribution of the perturbation of in the coefficient $\xi$, which is proportional to $\nu$ ($\delta\xi=\nu \xi \Delta$). The main message is that a viscous unified model with a constant viscosity parameter, $\xi=\xi_0$, remains a very competitive scenario for the dark sector. 

\section{The $\Lambda$vCDM model}

Although the unification scheme described in the last section appears as a viable alternative for the standard cosmology, the $\Lambda$CDM concept remains the baseline model for cosmology. Since the unified models are able to mimic (under some specific choices of the free parameters) the $\Lambda$CDM background expansion, only the perturbative analysis can distinguish between both scenarios. However, with a deeper analysis of the unified models, we realize that there is a fundamental difference, which is the value of $\Omega_{\rm vm0}$. For the unified models $\Omega^{uni}_{\rm vm0}\approx0.95$. A clear consequence of this, is that the epoch of matter-radiation equality is shifted to the past. In the standard cosmology, one has $\Omega^{std}_{\rm m0}\approx 0.3.$. This leads to a value $z^{std}_{eq}\sim3200$, for the redshift at which the universe becomes matter dominated. In the unified scenario, this moment is shifted by a factor $0.95/0.3 \sim 3.3$, i.e. $z^{uni}_{eq}\sim 10,000$. Hence, when dealing with such unified approach, one has to be very careful because some aspects of the pre-recombination physics is indeed modified. Then, a crucial observation that could definitely rule out the unified ideia is the redshfit of equality, i.e. this corresponds to a cosmological observable that tell us what is the correct value for $\Omega_{\rm m0}$. Since the peak of the matter power spectrum is sensitive to $z_{eq}$, large scale structure surveys can, in principle, constraint such value. Recently, the WiggleZ project has provided the first constraints on this quantity \cite{mreqWiggle}. Although the large uncertainty, their results are consistent with the standard value $z_{eq}\sim 3000$.  

There are also claims in the literature concerning measurements of galaxy clusters. Indeed, the baryonic matter fraction in X-ray luminous clusters provide compelling evidence that we live in a low density universe \cite{clustersOmegam}. If large galaxy cluster are reliable samples of the matter content of the universe, X-ray observations would be compatible with the standard model only for mean matter density values of order $\Omega_{\rm m0}\sim 0.2-0.3$ (see also \cite{Allen}). Hence, clusters seem to indicate that the unified scenarios fail in describing the correct abundance of matter in the universe.

Indeed it is difficult to find a alternative model that faces the standard $\Lambda$CDM universe. The most recent astronomical observations have preferred this model instead to put it under pressure. The PLANCK results are a recent example of this. Then, it seems a better strategy to find out how to solve the remaining problems of the $\Lambda$CDM than to propose a new baseline model for cosmology.

We will introduce now some of the problems of the standard cosmology that still need some explanation. A classical problem concerns the value of the cosmological constant \cite{LambdaProblem}, but we will focus on the the Cold Dark Matter component. Let us briefly describe them below:
\begin{arabiclist}
\item The missing satellite problem: In fact, this is a problem that arises from numerical simulations. The predicted number (calculated from the simulations) of small satellite (dwarf galaxies, for example) around a central galactic structure (like the Milk Way) is at least one order of magnitude larger than the observed one \cite{msp}. In other words, CDM forms too many structures at the sub-galactic level.  
\item The cusp-core problem: Together with the later problem, one considers the cusp-core problem a classical issue of the standard CDM paradigm. It is also a problem that has its origin in the numerical simulations. They point out that the internal density distribution ($\rho_g$) of galaxies follows a Navarro-Frank-White profile \cite{nfw}. For the inner part of galaxy, this profile predicts $\rho_g({r\rightarrow 0})\rightarrow \infty$, where $r$ is the radius of the galaxy. However, observations indicate that in fact the central region, let us say $r<1kpc$, is consistent with a core structure. Hence, the excess of clustering provided by the CDM model is not compatible with observations of the inner part of galaxies.  
\item The missing PLANCK clusters: Cosmology has entered in a new era with the recent data release of the Planck satellite. It is true that the main message obtained with Planck is that the $\Lambda$CDM model works very well. At the same time, it was, for some cosmologists, disappointing that no compelling evidence for new physics, e.g. dynamical behavior of dark energy, non-gaussianities, massive neutrinos, running of the inflationary spectral index has been found from the PLANCK-CMB analysis. However, the satellite also has the ability to identify and to count clusters through the signature of the Sunyaev–Zeldovich (SZ) effect. Using both the pure CMB data and the clusters count data, the Planck team has constrained the plan $\sigma_{8} x \Omega_{m0}$ for the standard $\Lambda$CDM model. It has been found a remarkable tension between both sources of data. The pure CMB data favour higher values for each parameter $\sigma_8$ and $\Omega_{m0}$ (see Fig. 11 in Ref. \cite{PlanckClusters}). This means that the Planck satellite has seen fewer cluster than expected. As concluded by the Planck team {\it ``This leads to a larger number of predicted clusters than actually observed''\cite{PlanckClusters}}. This result reinforces the previous discussion about the problems of the CDM scenario. We have now an extra evidence for the fact the standard CDM provides an excess of clustering and agglomeration. This happens not only for galactic structures but also at clusters scales.
\end{arabiclist}

The general ideia behind the problems listed above is the apparent excess of clustering predicted by CDM which is not observed at galactic-cluster scales. A proper inclusion of baryonic physics in the simulations can alleviate the small scale problems of the CDM paradigm. But, it does not solve the problems completely. 

Perharps, the correct theory for dark matter has to incorporate a new mechanism that is able to suppress the growth of CDM structures. As argued in Ref. \cite{viscousDM}, a viable scenario occurs if the structure of the $\Lambda$CDM model is preserved, but an almost vanishing bulk viscosity is associated to CDM only. This represents what we call the $\Lambda${\rm v}CDM model and the Hubble expansion in this case reads
\begin{equation}\label{HzviscousCDM}
H^2(a)=H^2_0 \left[\frac{\Omega_{\rm r0}}{a^4}+\frac{\Omega_{\rm b0}}{a^3}+\Omega_{\rm vm}(a)+\frac{\Omega_{\rm k}}{a^2}+\Omega_{\Lambda}\right].
\end{equation}

Note that we keep the cosmological constant. We allow CDM to have a negative pressure, without being the major cause of the accelerated expansion. The effects of the viscosity here are expected to be much smaller than in the previous unified case. This is basically for two reasons, namely, i) we have now $\Omega_{\rm vm}\sim 0.25$, while for the unified case $\Omega_{\rm vm}\sim 0.95$. Thus, the contribution of the viscous fluid to the total density is much smaller now. Also, ii) since the cosmological constant will drive the accelerated expansion, the viscosity of the fluid can be very small. Our goal now is to constrain the viscosity of dark matter. 

The viscous CDM energy density has to be determined by solving its conservation equation. As in the unified case, the viscous fluid has also a vanishing kinetic pressure but a small negative bulk viscous pressure (\ref{Pvis}).

We stick to the same choice for the bulk viscous coefficient (\ref{xi}) and hereafter assume a flat cosmology $\Omega_{k}=0$. With this choice
the energy-conservation equation for the viscous dark matter is written as
\begin{equation}
a \frac{d \Omega_{\rm vm}(a)}{da}+3\Omega_{\rm vm}\left(a\right)-\tilde{\xi}\left(\frac{\Omega_{\rm vm}\left(a\right)}{\Omega_{\rm vm0}}\right)^{\nu}\left[\frac{\Omega_{\rm r0}}{a^4}+\frac{\Omega_{\rm b0}}{a^3}+\Omega_{\rm vm}\left(a\right)+\Omega_{\Lambda}\right]^{1/2}=0,
\label{model2}
\end{equation}
where we have defined the dimensionless parameter
\begin{equation}
\tilde{\xi} = \frac{24\pi G \xi_0}{H_0}.
\end{equation}
As initial condition we set $\Omega_{\rm v}(a=1)=\Omega_{\rm v0}=0.3175$ \cite{Planck2013CP}. We solve this equation numerically.
Below, we will show results for the viscosity of dark matter in terms of the parameter $\tilde{\xi}$ and for this reason it is important the relate this quantity to the vCDM equation of state parameter today as
\begin{equation} \label{wv0}
w_{\rm vm 0}=-\frac{\tilde{\xi}}{3 \Omega_{\rm v 0}}.
\end{equation}
Note that CDM ($P=0$) is recovered if $\tilde{\xi}=0$.

The flat $\Lambda${\rm v}CDM cosmology has 4 free parameters, namely, $H_0, \Omega_{\rm vm0}, \tilde{\xi}, \nu$. Indeed, there a two more degrees of freedom than the standard $\Lambda$CDM. We can fix the exponent $\nu$ to the value $\nu=0$. This is only one of the possibilities, but let us concentrate on this case. Then, the remaining extra free parameter is $\tilde{\xi}$ which provides the viscosity of the dark matter fluid.

Our main goal here it to address the question, how viscosity can solve the CDM small scale problems. We will focus on the evolution of linear perturbations in the range of scales from dwarf galaxies to galaxy clusters. If the viscosity is able to produce some growth suppression, we are at least finding a clue for the solution of the above mentioned problems.

Starting with the perturbed metric (\ref{Newmetric}) and using the Einstein's equation and the energy and momentum conservations we obtain the following Meszaros-like equation for the sub horizon perturbations of the viscous CDM (see \cite{VeltenSchwarz11} for details). 
\begin{eqnarray}
a^{2}\frac{\,d^{2}\Delta_{\rm vm}}{da^{2}}+\left[\frac{a}{H}\frac{d
\,H}{da}+3+A(a)+B(a)k^{2}\right]a\frac{\,d\Delta_{\rm vm}}{da} \\ \nonumber
+\left[C(a)+D(a)k^{2}-\frac{3}{2}\right]\Delta_{\rm vm}=P(a),
\label{small}
\end{eqnarray}
\begin{eqnarray}
A(a)=-6w_{\rm vm}+\frac{a}{1+w_{\rm vm}}\frac{dw_{\rm vm}}{da}-\frac{2a}{1+2w_{\rm
vm}}\frac{dw_{\rm vm}}{da}+\frac{3w_{\rm vm}}{2(1+w_{\rm vm})}
\end{eqnarray}
\begin{eqnarray}
B(a)=-\frac{w_{\rm vm}}{3a^{2}H^{2}(1+w_{\rm vm})}\nonumber
\end{eqnarray}
\begin{eqnarray}
C(a)=\frac{3w_{\rm vm}}{2(1+w_{\rm vm})}-3w_{\rm vm}-9w^{2}_{\rm vm}-\frac{3w^{2}_{\rm
vm}}{1+w_{\rm vm}}\left(1+\frac{a}{H}\frac{dH}{da}\right) \\ \nonumber
-3a\left(\frac{1+2w_{\rm
vm}}{1+w_{\rm vm}}\right)\frac{dw_{\rm vm}}{da}+\frac{6aw_{\rm vm}}{1+2w_{\rm
vm}}\frac{dw_{\rm vm}}{da} \nonumber
\end{eqnarray}
\begin{eqnarray}
D(a)=\frac{w^{2}_{\rm vm}}{a^{2}H^{2}(1+w_{\rm vm})}
\end{eqnarray}
\begin{eqnarray}
&P&(a) = - 3 \nu w_{\rm vm} a \frac{d\Delta_{\rm vm}}{da} \\ 
&+& 3\nu w_{\rm vm} \Delta_{\rm vm} \left[-\frac{1}{2} +\frac{9w_{\rm vm}}{2} +\frac{-1-4w_{\rm
vm}+2w_{\rm vm}^2}{w_{\rm vm}(1+ w_{\rm vm})(1+2w_{\rm vm})} a \frac{d\,w_{\rm vm}}{da} -\frac{ k^{2}(1-w_{\rm vm})}{3H^{2}a^{2}(1+w_{\rm vm})}\right] \nonumber
\end{eqnarray}
where we have used the scale factor as dynamical parameter. The function $P(a)$ contains the terms proportional to the parameter $\nu$, i.e., the contributions from the perturbation of $\xi$ ($\delta\xi=\nu\xi\Delta$). Thus, for the case we are interested in ($\nu=0$) we have $P(a)=0$. A very interesting point here is that damping provided by the viscosity is scale-dependent. The scale-dependence appears not only as the standard contribution of the speed of sound, but it is also present in the friction term proportional to $\Delta^{\prime}$. If $w_{\rm vm}\neq0$, then $B(a)\neq 0$.

This equation has been used in \cite{VeltenSchwarz11} in the context of the viscous unified model. However, it is valid for any bulk viscous fluid. In practice, when compared with the unified case, only the background will be different. For the unified scenario it is required for the bulk viscous fluid a viscosity that is large enough to accelerate the current Universe. This leads to a substantial suppression of growth at small scales and the formation of small dark matter halos is challenged in the viscous unified cosmology. However, the unique case where the growth suppression is alleviated is the case $\nu=0$ \cite{VeltenSchwarz11}. This confirms that viable viscous models can be constructed for a constant bulk viscosity coefficient.

\begin{figure}[h!]
\centerline{\psfig{file=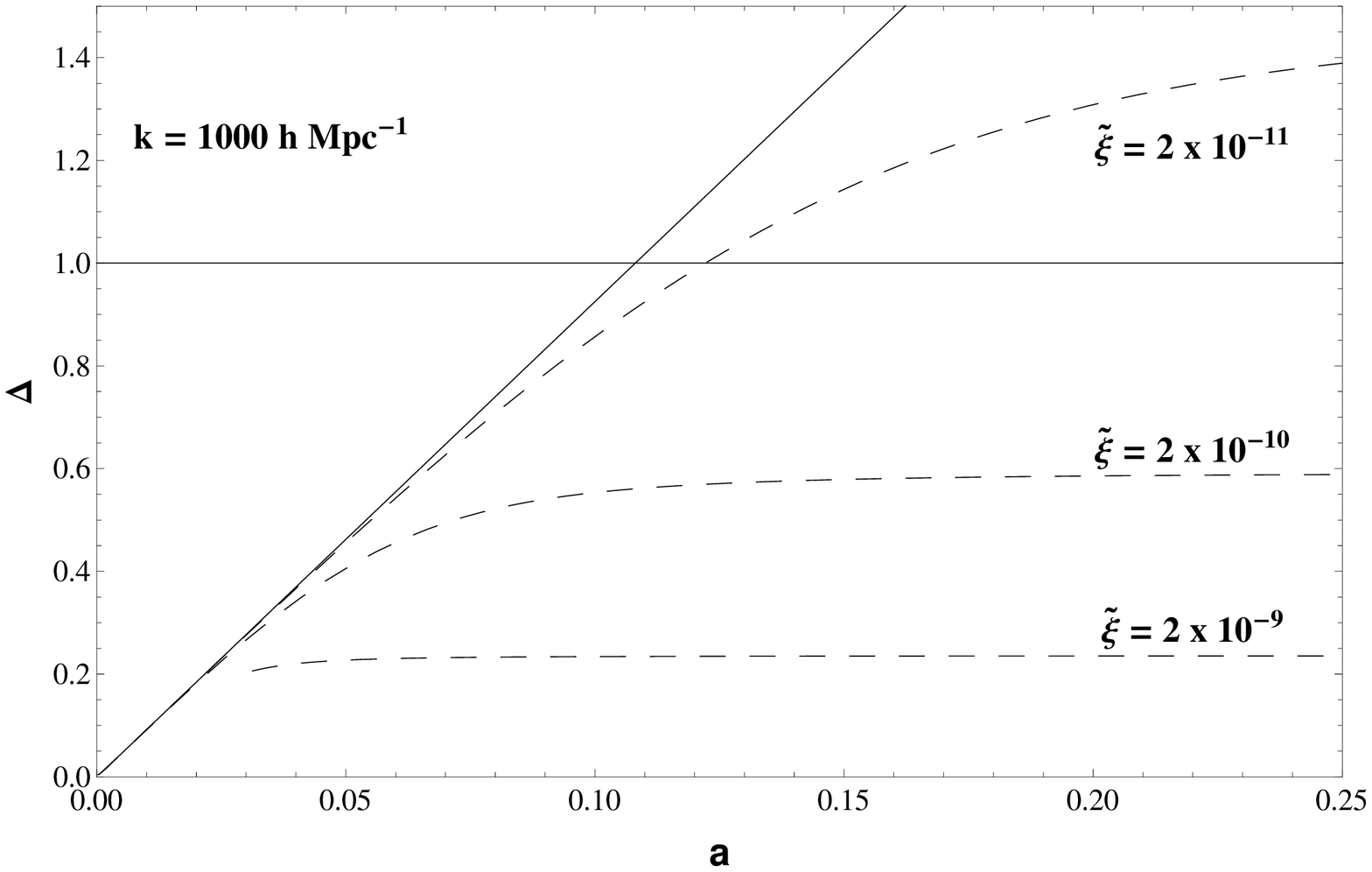,width=4.in}}
\centerline{\psfig{file=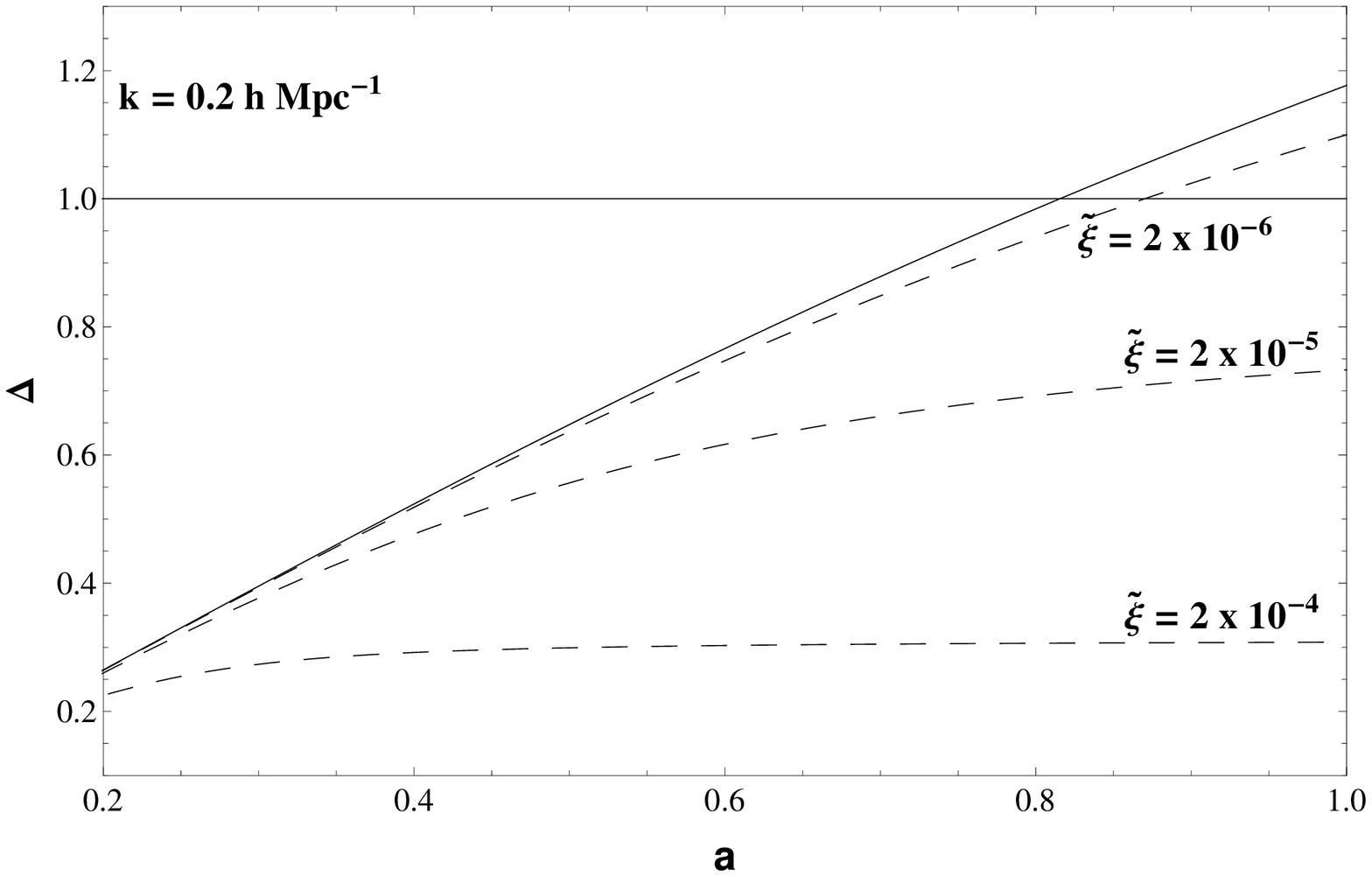,width=4.in}}
\vspace*{8pt}
\caption{Growth of Viscous dark matter halos. The density contrast $\Delta$ is plotted against the scale factor $a$. For both panels we have fixed $\nu=0$. The upper panel corresponds to a dwarf galaxy scale $k=1000 h {\rm Mpc}^{-1}$, while for the bottom panel we fixed $k=0.2 h {\rm Mpc}^{-1}$ which corresponds to a galaxy cluster scale. The solid horizontal line sets the nonlinear theory $\Delta=1$. Viscosity values are shown in each panel. \label{fig1}}
\end{figure}

Let us study dwarf galaxy and galaxy clusters scales. We show in Fig. \ref{fig1} the evolution of the density contrast $\Delta$ for both scales. Dwarf galaxy scale  ($k=1000 h {\rm Mpc}^{-1}$) in the upper panel and galaxy cluster scale ($k=0.2 h {\rm Mpc}^{-1}$) in the bottom panel. The scale factor is shown in horizontal axis. Remember that $a=1$ today. The horizontal line sets the onset of the nonlinear regime of the cosmological perturbation theory $\Delta=1$. The initial conditions were carefully calculated with help of the CAMB code \cite{CAMB}. In order to obtain them, we assume the $\Lambda$CDM model and calculate the amplitude of the dark matter perturbations at the matter-radiation equality $z_{eq}$. This is justified because for the values of the parameter $\tilde{\xi}$ shown in the plots, the effects of the viscosity are negligible at early times and thus we can safely assume that at $z_{eq}\sim 3000$ both models are the same. In fact, for these values of $\tilde{\xi}$, even the background dynamics of the $\Lambda${\rm v}CDM is the same as the standard $\Lambda$CDM. Remember the expression (\ref{wv0}) which says that the today's equation of state parameter of the viscous CDM is of the same order as $\tilde{\xi}$. Hence, we are working with almost negligible values for $w_{\rm vm0}$, but even so, the perturbations are sensitive to very small $\tilde{\xi}$ values.

The evolution of $\Delta$ is shown in the solid line corresponds to the standard CDM. It is basically a linear growth $\Delta\propto a$ until the moment at which the effects of the cosmological constant causes an almost imperceptible suppression. On the other hand, the dashed lines correspond to the growth of {\rm v}CDM halos for different values of $\tilde{\xi}$. Concerning dwarf galaxy scales, if the viscosity is $\tilde{\xi}=2$ x $10^{-10}$, the viscous structures would never reach the non linear regime. Of course, this is unacceptable since we do observe such structures in the universe. 

Comparing both panels in Fig. \ref{fig1} we note that the suppression is indeed scale dependent. For cluster scale, a value $\tilde{\xi}\sim 10^{-10}$ would be absolutely compatible with standard CDM. Only values of order $\tilde{\xi}\sim 10^{-5}$ would avoid the formation of structures like a galaxy cluster.

\section{Final Remarks}

We conclude this contribution commenting on specific topics.

\begin{itemize}

\item {\bf The $\Lambda${\rm v}CDM, a viable model?} It is important to remember that the $\Lambda${\rm v}CDM model represents a small correction to the standard cosmology. We are not proposing an alternative model. Instead, we are improving the standard picture via the inclusion of a physical mechanism that is very likely to occur in the universe. 

As shown in Ref. \cite{viscousDM}, background data place very weak constraints on the parameter $\tilde{\xi}$. For example, values of order $\tilde{\xi}\sim 0.1$ are already in agreement with Supernovae data, the Baryonic acoutic oscillations (BAO), the acoustic scale from CMB and an age for the universe of order $t_0 \approx 14$ Gyrs. 

Having the hierarchical structure formation scenario in mind, we know that the smallest structures form first. Since we observe structures like proto-galactic structures and dwarf galaxies today, we hav, at least, to guarantee the formation of the dark matter halos that host such structures. The analysis of the growth of viscous cold dark matter ({\rm v}CDM) halos shows that for values of order $\tilde{\xi}\sim 0.1$ structures would never form. In order to guarantee that dwarf galaxies scales reach to non-linear regime, i.e., $\Delta=1$, which is a necessary condition to form a virialized object, we set an upper bound on the allowed viscosity $\tilde{\xi}\lesssim 10^{-11} ~(10^{-3}$ Pa$\cdot$s in SI units). Such low $\tilde{\xi}$ values produce in practice the same background expansion as the $\Lambda$CDM model. Thus, it is impossible to distinguish the models using most of the available observational data. Numerical simulations would be required to predict the final clustering patterns.

\item {\bf Dark matter with negative equation of state?} In our approach dark matter has a negative pressure given by the viscous contribution $\Pi=-3H\xi$, where $\xi >0$. Of course, this happens because we set the kinetic pressure equals to zero $p_k=0$.

Let us now remember that DM particles have decoupled from the primordial plasma and have formed an isotropic gas in thermal equilibrium. From kinetic theory the pressure of a non-relativistic gas in this regime is given by
\begin{equation}
P=\frac{g}{3 h^3}\int \frac{p^2 c^2}{E} f(p)d^3 p\approx 4 \pi\frac{g}{3 h^3}\int \frac{p^4}{m_{\chi}} dp \rightarrow P= \rho c^2 \sigma^{2},
\end{equation}
where $g$ is the number of spin degrees of freedom, $h$ is the Planck constant, $p$ is the momentum of the particle that has energy $E=\sqrt{p^2+m^2 c^4}$ with distribution function $f$. For the velocity dispersion $\sigma^{2}=\left\langle \vec{v}^{2}\right\rangle/3c^{2}$ we assume a mean velocity square $\left\langle \vec{v}^{2}\right\rangle=81 \times 10^{14} cm^2/s^2$, leading to $\sigma^{2}=3\times 10^{-6}$. It is, of course, a negligible number. However, note that $\tilde{\xi}\sim 10^{-11}$ produces a today's equation of state parameter for our {\rm v}CDM ({\ref{wv0}}) that is much smaller that $w \sim 10^{-6}$. Thus, this proves that the viscosity has to be seens as small deviation from the CDm paradigm. 

Therefore, since the viability of the $\Lambda${\rm v}CDM model is conditioned to values of order $\tilde{\xi}\sim 10^{-11}$, the inclusion of the the kinetic pressure of order $w_{\rm dm}\sim 10^{-6}$ guarantee that the total (or effective) pressure of the viscous dark matter remains positive. 

On the other hand, it would interesting to investigated what are the impact of a bulk viscous pressure on warm dark matter models which have a small (but non negligible) positive pressure. 

\item {\bf What is the correct form for the coefficient $\xi$?} The choice (\ref{xi}) for the coefficient of bulk viscosity is quite phenomenological. 

The transport coefficients are calculated in kinetic theory as powers of the temperature $\xi\equiv\xi(T)$. Using the appropriate thermodynamical relations we can replace the temperature $T$ by the density of the fluid. However, it is usual in the literature to use the coefficient of bulk viscosity as a function of the background expansion $\xi\equiv\xi(H)$. Of course, this is valid only for a one-fluid description of the cosmic medium, where the bulk viscous fluid dominated the dynamics the therefore $\rho_{\rm v}\propto H^2$. In our case, the ${\rm v}$CDM coexists with other components. Then, if we set a dependence like $\xi_{\rm vCDM}=\xi_{\rm vCDM}(H)$ it would be necessary to justify the coupling between ${\rm v}$CDM and the other fluids, i.e., why does the dark matter viscosity depends on the baryonic matter, radiation and dark energy properties? Thus, the choice (\ref{xi}) seems to be the most adequate.

An interesting approach for the viscous dark matter idea could be the ''dark goo`` model as proposed in Ref. \cite{darkgoo}, where it is used a scalar field representation for the viscous fluid and the proper bulk viscosity for scalar theories has been adopted.

\end{itemize}
\section*{Acknowledgments}

I wish to thank the organizers of the 49th Winter School of Theoretical Physics ``Cosmology and non-equilibrium statistical mechanics''. It is a pleasure to thank Diego Pav\'on, Dominik Schwarz, J\'ulio Fabris, Gilberto Kremer and Winfried Zimdahl for helpful discussions. This work was financially supported by CNPq (Brazil). The author is also thankful to the Department of Physics of Bielefeld University.

\end{document}